\documentstyle[aps,prl,epsf]{revtex}  

\begin{document}
\twocolumn[\hsize\textwidth\columnwidth\hsize\csname
@twocolumnfalse\endcsname

\title{The buckling transition of 2D elastic honeycombs: Numerical simulation and Landau theory}
\author{E. A. Jagla}
\address{The Abdus Salam International Centre for Theoretical Physics\\
Strada Costiera 11, (34014) Trieste, Italy}
\maketitle

\begin{abstract}
I study the buckling transition under compression of a two-dimensional, hexagonal, regular elastic 
honeycomb. Under isotropic compression,  the system
buckles to a configuration consisting of a 
unit cell containing four of the original hexagons. This buckling pattern
preserves the sixfold rotational symmetry of the original lattice but is chiral, and can be
described as a combination of three different elemental distortions in directions
rotated $2\pi/3$ from each other. Non-isotropic compression may induce
patterns consisting in a single elemental distortion or a superposition of two 
of them. The numerical results compare very well with the outcome of a Landau theory 
of second order phase transitions.

\end{abstract}
\vskip2pc] \narrowtext


\section{Introduction}
A two dimensional honeycomb structure formed by solid walls is the prototype
of a cellular solid \cite{ga1}. These are materials widely used in
applications due to their remarkable mechanical properties, for instance its
capacity for energy absorption under impact, and its low weight. Energy absorption
is related to plastic deformation under stress. But still the ideally elastic
and perfectly uniform two dimensional honeycomb presents some not 
completely solved puzzles. Under
compressive stress, it has a buckling transition in which
some (or all) of their walls bend. This transition is reminiscent of the well known
buckling transition of an elastic bar under compressive stress at its extremes\cite{timo}.
There has been some controversy on what the buckling mode of a regular honeycomb should be.
On one hand, in their original
work\cite{ga2}, Gibson and Ashby presented the results of an experiment using 
an elastomeric honeycomb, under what they called biaxial loading, 
in which they observed a non-trivial buckling pattern
consistent with a symmetry breaking in which four original cells form the new 
repetitive motif of the material. In a posterior paper\cite{hw1}, Hutzler and Weaire performed
numerical simulations and did not observe this pattern, but instead a
buckling mode equivalent to that obtained under uniaxial stress. They argue that the
pattern observed in Ref. [3] was a consequence of finite size effect, and the use of
flat confining walls. Numerical results taking into account these
effects\cite{hw2} did show the pattern observed by Gibson and Ashby. 
Very recently, Okomura, Ohno and 
Noguchi\cite{japoneses} have studied the problem using a combination 
of a homogenization technique and finite elements numerical simulations. 
Their results do not agree with those of 
Hutzler and Weaire\cite{hw1}. Instead, they found buckling patterns that can be interpreted as 
a superposition of three individual buckling modes.
They also found that whether one, two, or three of these modes are 
active depends on the degree of anisotropy of
the externally applied strain.

In view of the aforementioned contradiction between \cite{hw1} and \cite{japoneses}, and considering that 
the techniques employed in both cases are quite different, an independent investigation
to determine which of the
two results is correct seems appropriate. 
In the first (numerical) part of this paper I will show that 
appropriately done numerical simulations using the 
technique used in \cite{hw1} do not support the
results claimed there, but instead those reported in \cite{japoneses}. 
In the second (more theoretical) part 
I will show how the results obtained in the simulations 
are fully compatible with the predictions of a
Laundau theory of second order phase transition applied to the buckling problem.
This theory allows to obtain at once
the buckled configuration of the system under a generic form of the macroscopically 
homogeneous applied deformation. 

\section{Numerical Simulation}

I have simulated a two dimensional honeycomb through the technique used in Refs. \cite{hw1}, \cite{hw2},
namely by considering the honeycomb walls as one
dimensional rods, and including stretching and bending energy as 
\begin{eqnarray}
E_{stretch}&=&\frac 1 2 k_s
\int\left( \frac {dl}{dl_0} -1 \right )^2 dl_0\nonumber\\
E_{bend}&=&\frac 1 2 k_b
\int c^2 dl_0
\end{eqnarray}
where $c$ is the local curvature.
To discretize these expressions I have used 7 intermediate points between any two neighbor vertices, 
but particular cases where checked using
18 points, to guarantee the absence
of noticeable effects due to discretization. The only essential parameter of the model is
$k_b/L^2k_s$, where $L$ is the length of the individual rods. 
This ratio is physically related to the fraction $\Lambda$ of two dimensional space that is occupied
by the rods \cite{ga1,hw1}, namely $\Lambda=4\sqrt{k_b/L^2k_s}$.
The simulations presented below were done at $k_b/L^2k_s=4.5\times 10^{-4}$ (then $\Lambda=0.085$), 
but additional checks indicate that
the results are not qualitatively dependent on the precise value of the parameter.
All previously obtained buckling patterns for perfect honeycombs can be accommodated within a
$2\times 4$ unit cell. Then the elemental cell I simulated is precisely the 
$2\times 4$ cell shown in Fig. 1(a), with periodic boundary conditions. The simulation method consists
in calculating the forces acting on all points of the discretized system, and updating their positions
using a viscous dynamics.
The control variable was the macroscopic strain, that can be changed varying the size 
and shape of the simulation box.
Stresses in the system can be
evaluated both by numerical differentiation of the total energy with respect to strain, 
and by direct summation in terms of the forces between particles. The equivalence of the 
two results allows to check for consistency and convergence of the simulation.

Before indicating the results obtained, it is clarifying to discuss qualitatively the
behavior observed (see \cite{japoneses}). The buckling structures that appear are related 
to reaccommodation of the vertices of the honeycomb structure, in such a
way that lines of vertices forming zig-zag chains shift relatively to neighbor chains, as qualitatively
indicated in Fig. 1(b). There are three of these modes, that will 
be refereed to as the elementary modes of buckling. The patterns they generate will be called the uniaxial
patterns. They are characterized by
the unitary vectors shown in Fig. 1(e).
Whether one, two, or the three elementary modes acquire non-zero amplitude at the buckling
transition depends on the macroscopic strain applied.
The qualitative pictures of Fig. 1(c) and (d)
show the effect of combining two or three elementary modes. We see that the patterns of Fig. 1(b) and (c) are
identical (except for the wall bending, that in this qualitative picture is not taken into account) 
to the patterns in Ref. [3]. The pattern in Fig. 1(c) will be referred to as the Gibson-Ashby pattern.
The pattern in Fig. 1(d), that combines the three elementary modes, 
preserves the hexagonal symmetry of the structure 
(this will be referred to as the symmetric pattern).
A buckling mode of this symmetry has been observed and simulated in \cite{papa}, but only in the
case of {\em plastic} buckling.
To my knowledge, the only prediction of this mode for a perfectly elastic honeycomb is contained in
\cite{japoneses}.
It is remarkable that this pattern has lost
the mirror symmetry plane of the original structure: 
it is a chiral pattern. The chirality can be defined as the sign of 
the product of the
amplitudes of the three elementary modes from which the pattern is constructed.

In the numerical simulations, starting from the unstrained structure and
isotropically compressing it, I have observed that the symmetric pattern is the stable
configuration after buckling.  
However, it has to be mentioned that the transition from the unbuckled to the symmetrically buckled
configuration, although continuous (then implying no surmounting of any energy barrier)
may take a long computational time, and that other patterns (which correspond to saddles of the energy) can be
transitory observed. 
For instance, I show in Fig. 2 the mechanical energy of the
system as a function of the compressive strain $s$ in a single run increasing $s$ at a constant rate. 
At the buckling point (at $s\simeq 0.0037$, note that I define $s$ as $s=(s_x+s_y)/2$), and in the
particular run shown, the system seems to buckle to the Gibson-Ashby 
pattern shown in Fig 3(b). This is not a
minimum but a saddle of the energy, however it lasts long enough, and 
the energy of this branch can be numerically followed (as seen in
Fig. 2 with open triangles) up to some strain at which it is observed 
to transform to the real energy minimum which
corresponds to the symmetric pattern. In fact, when preparing the system in the symmetric pattern for
large strain, and upon reduction of the strain, we obtain the results indicated by full circles in Fig. 2, which
correspond to the real energy minimum of the system. In addition, if the system is prepared 
in the uniaxial or in the Gibson-Ashby pattern at
large strain, and strain is reduced (sufficiently rapidly for the configuration not
to destabilize, and sufficiently slowly so as energy can be calculated accurately) we can follow
the energy of these two configurations, as indicated by the stars, and the full triangles in Fig. 2.

The uniaxial and Gibson-Ashby patterns may become the stable buckling modes under appropriate 
non-isotropic loading. For instance, by compressing along the $x$ ($y$) direction, and keeping the perpendicular
direction unstrained, I have observed the Gibson-Ashby (uniaxial) pattern  to be stable after
buckling.
The stability of the Gibson Ashby pattern under particular uniaxial loading 
agrees again with the results in \cite{japoneses}, but not with those claimed in \cite{hw1}.
We will see now how all these results can be fully systematized within a theoretical framework.

\section{Landau theory of the buckling transition}

The results presented in the previous section are enough
numerical input to construct the Landau theory of this peculiar second order transition.
In fact, we see in Fig. 2 that the energies of all buckled configurations 
(whether the true minimum or the saddles) meet together in value and derivative at
the buckling point, coinciding also at that point in value and derivative with the branch corresponding
to the unbuckled system. This means that at the buckling point, in a generic parameter space, the state point of
the system passes from a configuration with a single minimum (for the unbuckled state) to one with different
minima and saddles in a continuous manner.

I will present a Landau description in which
the order parameters are the three (small) amplitudes $\phi_1$, $\phi_2$, $\phi_3$ of the 
elementary modes. For convenience, these three modes will be associated with the unitary vectors
${\bf v}^1$, ${\bf v}^2$, and ${\bf v}^3$ shown in Fig. 1(e).
The free energy of the system (in the present case it actually corresponds simply to the elastic energy)
must a scalar, and then it can only contain combinations of the amplitudes and the
(eventually anisotropic) external strains that
are invariant with respect to the symmetry operations of the lattice.
Considering for the moment only the isotropically compressed case, 
we should look for invariant combinations of the amplitudes.
Up to fourth order those available are:
\begin{eqnarray}
{\phi_1}^2+{\phi_2}^2+{\phi_3}^2\nonumber\\
\left({\phi_1}^2+{\phi_2}^2+{\phi_3}^2\right)^2\\
\phi_1^2\phi_2^2+\phi_2^2\phi_3^2+\phi_3^2\phi_1^2\nonumber
\end{eqnarray}
Then in this case the most general form of the
free energy describing a second order transition is
\begin{eqnarray}
F= \alpha(s_c-s)\left(\phi_1^2+\phi_2^2+\phi_3^2\right)+\beta \left (\phi_1^2+\phi_2^2+\phi_3^2 \right)^2 +\nonumber\\
-\gamma \left(\phi_1^2\phi_2^2+\phi_2^2\phi_3^2+\phi_3^2\phi_1^2\right)
\label{uno}
\end{eqnarray}
with numerical constants $\alpha$, $\beta$, and $\gamma$. This free energy has a single minimum at $\phi_p=0$ ($p=1,2,3$) for $s<s_c$, representing the unbuckled state.
For $s>s_c$, this expression has saddles (or minima) at the following values of the amplitudes:

\begin{eqnarray}
\phi_p^2&=& A, ~~\phi_q=0,~~ \phi_r=0\\
\phi_p^2&=& B, ~~\phi_q=B,~~ \phi_r=0\\
\phi_p^2&=& C, ~~\phi_q=C,~~ \phi_r=C
\end{eqnarray}
where $p$, $q$, $r$ are arbitrary permutations of 1, 2, 3,
and $A$, $B$, $C$, are given by:
\begin{eqnarray}
A&=& \frac{(s-s_c)\alpha}{2\beta}\\
B&=& \frac{(s-s_c)\alpha}{4\beta-\gamma}\\
C&=& \frac{(s-s_c)\alpha}{6\beta-2\gamma}
\end{eqnarray}
They correspond respectively to the uniaxial, Gibson-Ashby, and symmetric patterns.
The corresponding values of the free energy are,
\begin{eqnarray}
f_{uni}-f_{unb}&=&-\frac{(s_c-s)^2\alpha^2}{4\beta}\\
f_{GA}-f_{unb}&=&-\frac{(s_c-s)^2\alpha^2}{4\beta-\gamma}\\
f_{sym}-f_{unb}&=&-\frac{(s_c-s)^2\alpha^2}{4\beta-4\gamma/3}
\end{eqnarray}
(I have subtracted the free energy of the unbuckled system $f_{unb}$, that is taken as 
zero in the Landau theory, but that should be included when comparing with the results 
of the numerical simulations).
We see that the symmetric
pattern is the minimum energy one for $\gamma>0$ (whereas the uniaxial pattern 
provides the absolute minimum if $\gamma<0$).
Then, in order to agree with the simulation results, we will assume $\gamma>0$.
The physical justification for the positive sign of $\gamma$ 
is not provided by the Landau theory, and should come from an explicit evaluation of the
mechanical energy of the honeycomb.
From the previous values of the free
energy a parameter free relation can be obtained and compared with the numerical results. 
First note that he ratio $\beta/\gamma$ can be obtained for instance as
\begin{equation}
\frac {\gamma}{\beta}=3\left (\frac{f_{sym}-f_{uni}}{f_{sym}-f_{unb}}\right)
\end{equation}
This in particular should be independent of the actual value of the compression $s$ (as long as $s>s_c$). 
From the numerical results it can be obtained that this ratio is in fact 
independent on $s$, and the numerical value
is approximately ${\gamma}/{\beta}\simeq 0.01$ for the parameters of the
simulation. Then note that to a very good approximation we have 
\begin{equation}
\frac{f_{GA}-f_{sym}}{f_{uni}-f_{sym}}=\frac13
\end{equation}
This is a parameter free relation that has to be satisfied in our numerical simulations. From the data in Fig. 2,
it can be in fact verified that this is very accurately satisfied. This is a strong evidence that the
present Landau theory describes the physics of the present buckling transition.

In order to make the theory more complete, I want to consider now the possibility of non-isotropic external
loading on the system. This means that instead of a single parameter $s$, we have now a generic (symmetric)
strain tensor $s_{ij}$ ($i,j=1,2$) applied onto the system (the previously introduced 
isotropic compression $s$ is related to the trace of this tensor). 
This has to be introduced into the free energy in a symmetrically
invariant form. To lowest order I will include it only in the second order term, which is the one
that triggers the transition. Using the
unitary vectors ${\bf v}^1$, ${\bf v}^2$, ${\bf vu}^3$ of the elementary modes, two different terms
quadratic in the amplitudes can be written:
\begin{eqnarray}
F^{(1)}&\sim&\sum_{i,j=1,2}\sum_{p,q=1,2,3}s_{ii}v^p_iv^q_i a_{pq} \phi_p\phi_q\\
F^{(2)}&\sim&\sum_{i,j=1,2}\sum_{p,q=1,2,3}s_{ij}v^p_iv^q_j b_{pq} \phi_p\phi_q
\end{eqnarray}
where $a_{pq}$ and $b_{pq}$ are arbitrary numeric matrices. However, these expressions have
to be invariant under permutation of the elementary vectors (since this is a symmetry operation obtained
by the mirror symmetry along the line containing the third vector) and sign change of any of the amplitudes
(which is obtained by a particular spatial translation allowed by symmetry). Then it is obtained that  
both $a_{pq}$ and $b_{pq}$ matrices should be proportional to the identity, i.e,
\begin{eqnarray}
F^{(1)}&\sim&\sum_{i,j=1,2}\sum_{p=1,2,3}s_{ii}v^p_iv^p_i \phi_p^2\\
F^{(2)}&\sim&\sum_{i,j=1,2}\sum_{p=1,2,3}s_{ij}v^p_iv^p_j \phi_p^2
\end{eqnarray}

Then $F^{(1)}$ becomes proportional to $s_{ii}(\phi_1^2+\phi_2^2+\phi_3^2)$ and is the term
considered in the isotropically compressed case.

To analyze the second contribution it can be more convenient to use the following 
definition of the three independent components of the strain tensor:

\begin{eqnarray}
s&=&(s_{11}+s_{22})/2\nonumber\\
s_2&=&(s_{11}-s_{22})/2\nonumber\\
s_3&=&s_{12}=s_{21}\nonumber
\end{eqnarray}
which represent the applied deformation in a more physical way: $s$ represents an 
isotropic compression (we already used this), whereas
$s_2$ and $s_3$ are the two independent shear modes, which are related by a $\pi/4$ rotation.
In terms of these variables, and using explicitly the components of the unitary vectors
we finally arrive to the following form of the free energy:

\begin{eqnarray}
F&=&
\alpha\left[(s_c-s)\left(\phi_1^2+\phi_2^2+\phi_3^2\right)\right]+\nonumber\\
&+&\delta\left[
s_2\left(\phi_1^2-\phi_2^2/2-\phi_3^2/2\right)+s_3\frac{\sqrt3}{2}\left(\phi_3^2-\phi_2^2\right)\right]+\nonumber\\
&+&\beta \left (\phi_1^2+\phi_2^2+\phi_3^2\right )^2 
-\gamma (\phi_1^2\phi_2^2+\phi_2^2\phi_3^2+\phi_3^2\phi_1^2)
\label{kb}
\end{eqnarray}
This is the final expression for the free energy close to the buckling transition.
Minimizing it we can obtain the buckled state under any particular combination
of the three independent strains $s$, $s_2$, and $s_3$.
Note that as this expression has to describe also the unstrained system, and then
$\alpha s_c$ and $\delta$ will be respectively proportional to bulk and shear modulus of the
original structure, then typically $\delta<<\alpha s_c$.

I want to describe now the buckling mode map of the system, namely, what the amplitudes of the
three elementary modes are for any choice of the strain tensor.
First notice the following scaling of the free energy: If we consider 
the values of the three order parameters at the minimum of (\ref{kb}), namely $\phi^p_{min}$, 
to be a function of
$s_c-s$, $s_2$, and $s_3$, then the following relation is satisfied:
\begin{equation}
\phi^p_{min}(s_c-s,s_2,s_3)=\lambda^{-1/2}\phi^p_{min}(\lambda(s_c-s),\lambda s_2,\lambda s_3)
\end{equation}
This implies in particular that the borders between different regions in the parameters
space $s_c-s$, $s_2$, and $s_3$ are spanned by rays propagating from the origin.

I will analyze a couple of particular cases.
First consider the case $s_3=0$, i.e., purely compressive strains along $x$ and $y$ (although non
necessarily equal).
I show in Fig. 4(a) the map of buckling modes in the $s$-$s_2$ plane for 
this case. The borders between different regions can be worked out analytically.
All of them are straight lines emanating from the point $s=s_c$, $s_2=0$, as the previous argument
indicates.
The transition between unbuckled ($\phi_p\equiv0$) and
uniaxial pattern ($\phi_1\ne0$) can be easily obtained setting $\phi_2=\phi_3=0$
in (\ref{kb}). The limit line is given by $s_2=(s-s_c)\alpha/\delta$. The transition 
line between the unbuckled and the Gibson-Ashby pattern ($\phi_2=\phi_3\ne0$) is obtained along the same
lines, as $s_2=-2(s-s_c)\alpha/\delta$. Increasing $s$ at $s_2=0$, the symmetric
pattern appears at $s=s_c$, as we already know from the isotropically compressed case. 
The symmetric pattern looses its strict rotational symmetry for any $s_2\ne0$.
However, it still has the three elementary model active as long as we are within
the V-shaped region in Fig. 4(a). The limits of this region
are given by 
\begin{equation}
s_2^{(uni \rightarrow sym)}=\frac13\frac{\alpha\gamma}{\beta\delta}\frac{s_c-s}{\left(1-\frac{\gamma}{3\beta}\right)}
\end{equation}
for the transition
to the uniaxial pattern, and 
\begin{equation}
s_2^{(GA \rightarrow
sym)}=-\frac16\frac{\alpha\gamma}{\beta\delta}\frac{s_c-s}{\left(1-\frac{\gamma}{3\beta}\right)}
\end{equation}
for the transition to the Gibson-Ashby pattern. Note the exact relation
\begin{equation}
s_2^{(GA \rightarrow sym)}/s_2^{(uni \rightarrow sym)}=-2
\label{dos}
\end{equation}
valid for any values of the parameters of the free energy. 

The results of Fig. 4(a) 
are fully compatible with the results in \cite{japoneses} (in particular, relation (\ref{dos}) is
very well satisfied). We note that for
the present parameters the stability of the single elementary mode and Gibson-Ashby pattern 
that was obtain numerically under uniaxial compression is recovered. 

As an additional example I show the map of buckling modes
in the $s_2$-$s_3$ plane for some $s>s_c$ in
Fig. 4(b). 
Note the nice symmetry of this pattern, which has one two or three elementary modes active 
depending on the particular choice of the applied strains $s_2$ and $s_3$ (remember 
that $s_2$ and $s_3$ are related by a rotation of $\pi/4$).
Again, all borders between different sectors are straight lines. The analytical expression
for the line separating sectors 1 and 1,2 is given by 
\begin{equation}
s_3=\sqrt 3 s_2 \left (1-\frac{\gamma}{3\beta}\right)-\frac{\alpha\gamma}{\sqrt3\beta\delta}(s_c-s)
\end{equation}
All other lines can be obtained then 
from symmetry.

To finish, we note that all transitions in the parameter space are continuous, namely, there
are no jumps of the order parameters at any point, and there is no possibility of metastabilities
either.


\section{Conclusions}

The buckling mode of an elastic two dimensional honeycomb 
provides an example of non-trivial patterns with
symmetry breaking appearing in a very simple mechanical 
system. Remarkably, for isotropic compression the symmetry 
breaking produces the appearance of a chiral ground state.
This problem is also a realization of a second order transition 
that can be accurately modeled through a Landau theory
constructed on the basis of the symmetry of the problem.
The agreement between the Landau theory and the numerical 
simulation is seen to be very good.

\begin{figure}
\epsfxsize=3.3truein
\vbox{\hskip 0.05truein
\epsffile{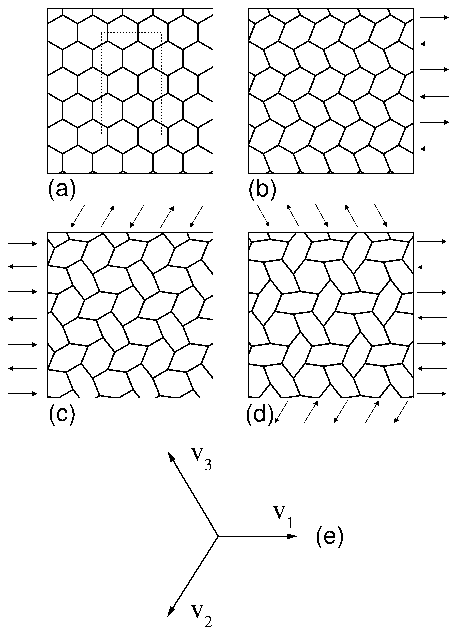}}
\medskip
\caption{(a) The hexagonal starting lattice. Dotted box is the system actually simulated.
(b) Upon shifts of the vertices as indicated by the arrows, the uniaxial pattern is obtained. Note 
however that there are three equivalent ways of generating this pattern, that can be characterized by the unitary
vectors shown in (e). Combining two or the three
of them we obtain the configurations in (c) and (d).}
\label{f1}
\end{figure}

\begin{figure}
\epsfxsize=3.3truein
\vbox{\hskip 0.05truein
\epsffile{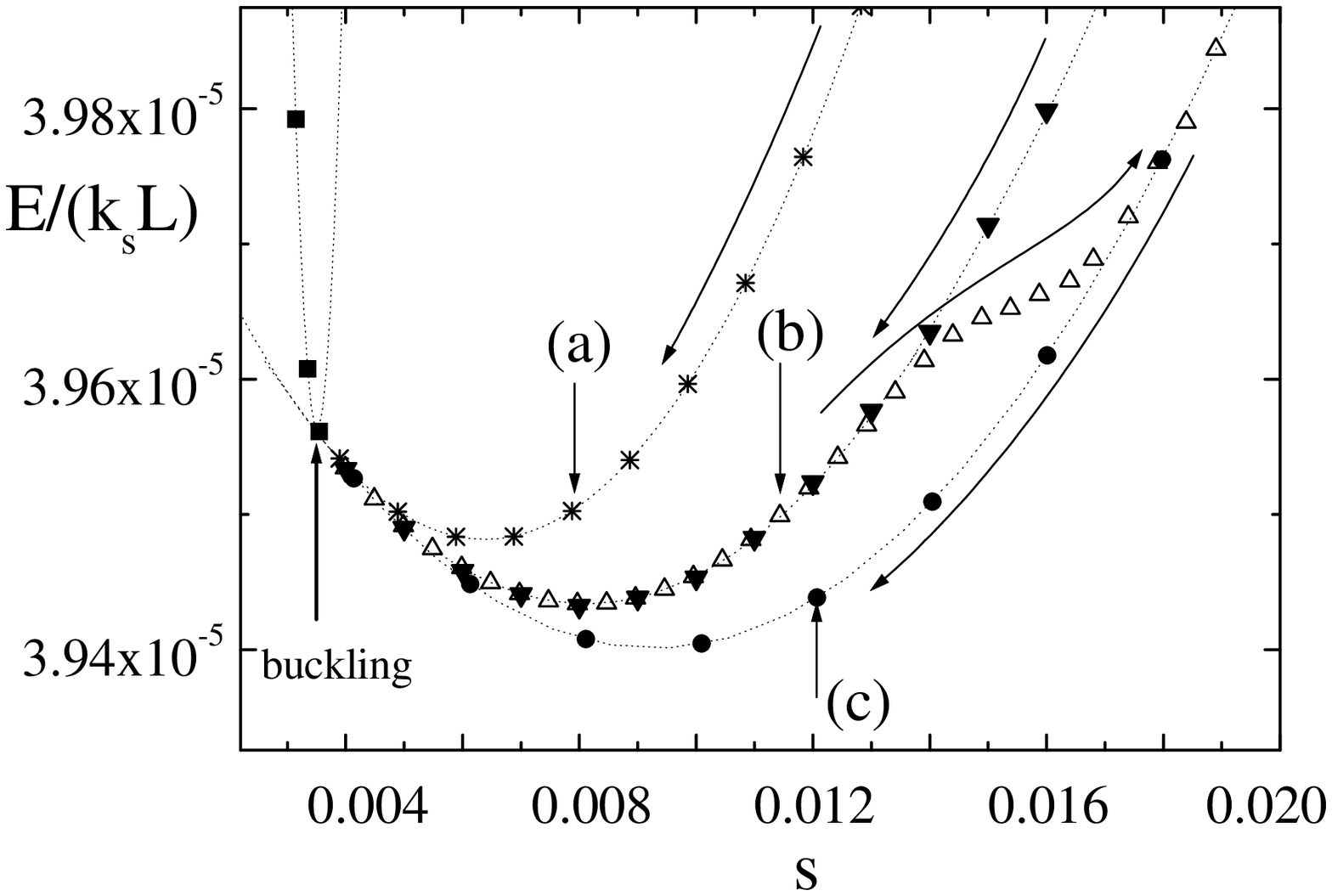}}
\medskip
\caption{Total energy of the simulated system (dotted box in Fig. 1(a)) 
as a function of compressive strain $s\equiv (s_x+s_y)/2$. 
A global, linear contribution in $s$
has been subtracted to
appreciate tiny differences in energy. All continuous lines are quadratic fitting of the date close 
to the buckling point. 
Results are shown from a running from the unbuckled state (squares), in which a Gibson-Ashby pattern 
is generated first (open triangles). This is however a saddle of the energy, and is seen to
transform to the symmetric pattern after some time. 
The other three curves show the results 
starting from the uniaxial (stars), Gibson-Ashby (full triangles), 
and symmetric (circles) patterns at high strain, 
and reducing it towards the unstrained configuration.
Note how the minimum energy of the buckled state is obtained for the symmetric pattern. The
letters indicate where the snapshots in Fig. 3 were taken.}
\label{f3}
\end{figure}

\begin{figure}
\epsfxsize=3.3truein
\vbox{\hskip 0.05truein
\epsffile{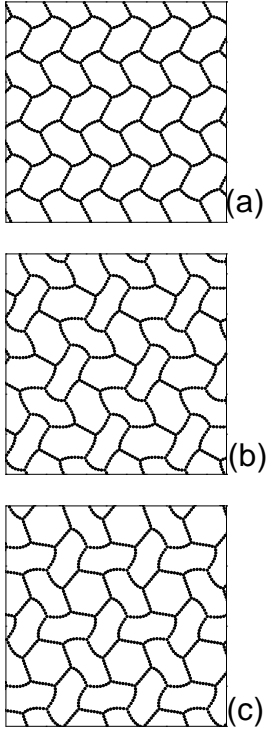}}
\medskip
\caption{The uniaxial, Gibson-Ashby, and symmetric buckling modes. The snapshots correspond to the points
indicated in Fig. 2. For isotropic compression, the symmetric pattern provides the minimum energy, the other two correspond to saddles, and
eventually destabilize. However, they can be made stable under appropriate non-isotropic loading
(the displacements with respect to the hexagonal configuration have been amplified by a factor of
5 to render the geometrical structure more visible).}
\label{f2}
\end{figure}

\begin{figure}
\epsfxsize=2.5truein
\vbox{\hskip 0.05truein
\epsffile{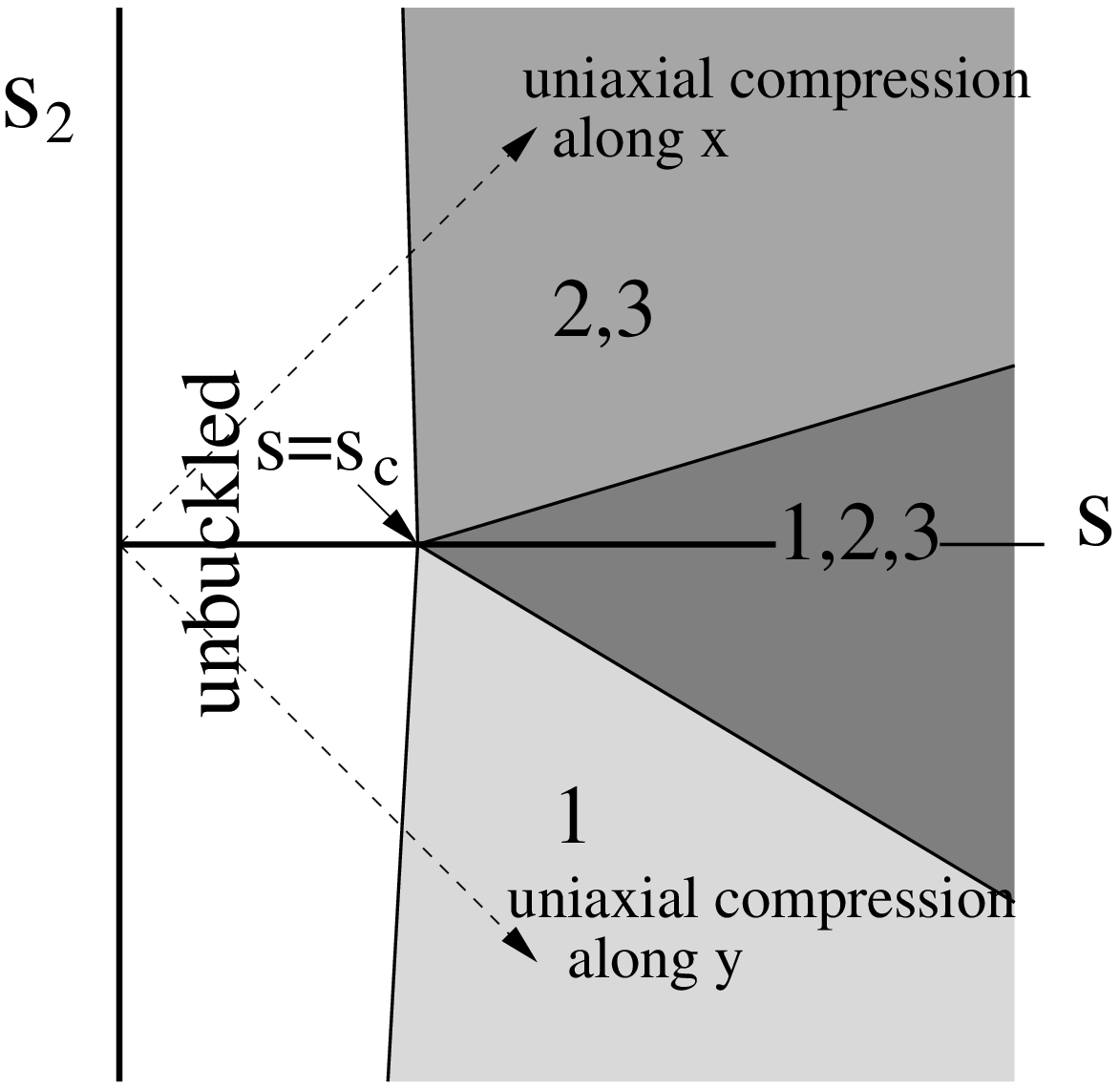}}
\medskip
\epsfxsize=2.5truein
\vbox{\hskip 0.05truein
\epsffile{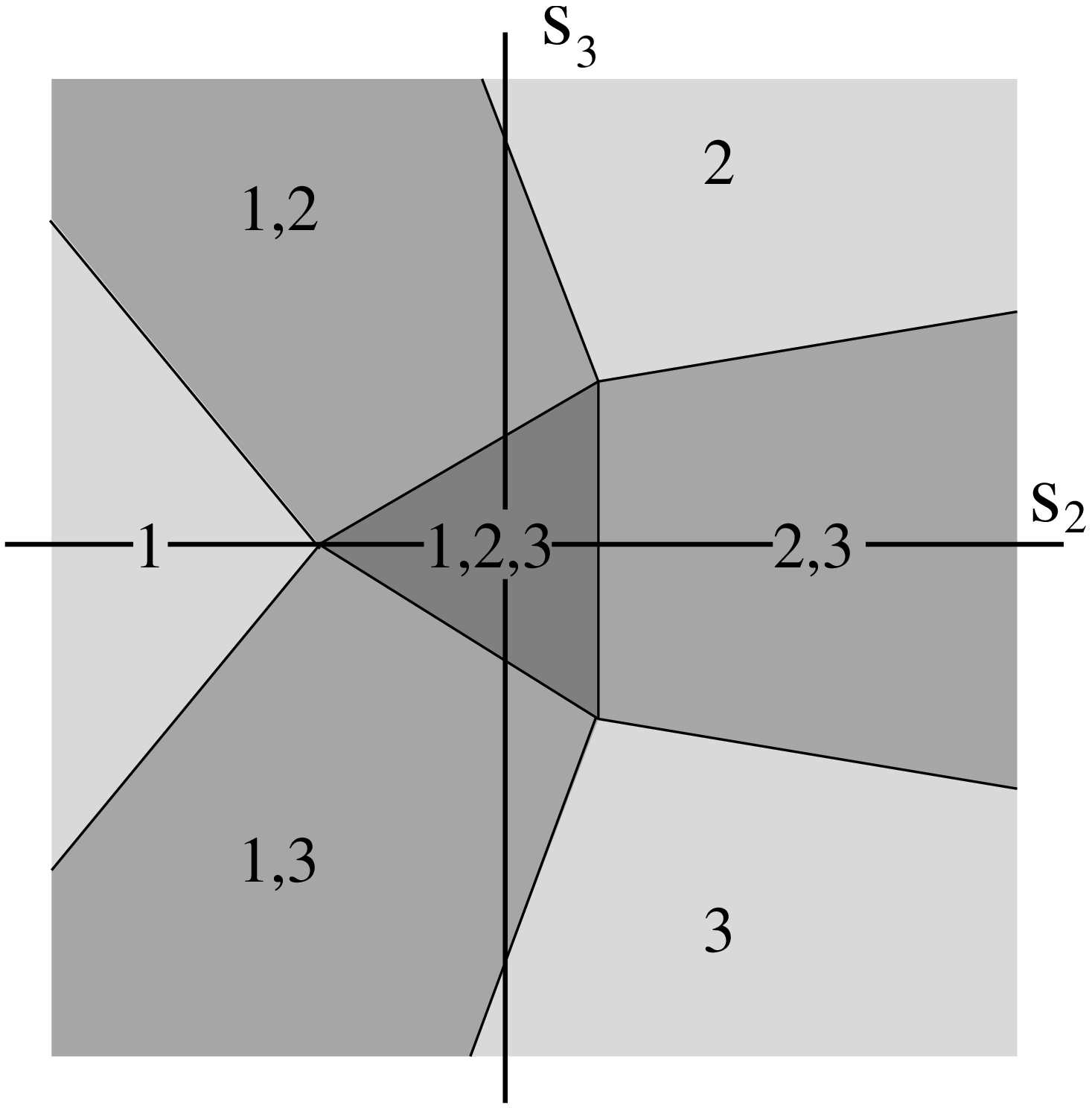}}
\medskip
\caption{The buckling modes map in the $s$-$s_2$ plane for $s_3=0$
(a), and in the $s_2$-$s_3$ plane for a constant value of $s>s_c$
(b). In each region, the numbers indicate which elementary modes are active (see Fig. 1).
The analytical expressions for the limits between different regions are given in the text.}
\label{f4}
\end{figure}

\end{document}